\documentclass[prb,twocolumn,superscriptaddress]{revtex4}
\usepackage{graphicx}
\renewcommand{\narrowtext}{\begin{multicols}{2} \global\columnwidth20.5pc}
\renewcommand{\v}[1]{{\bf #1}}
\newcommand{\sign}{{\rm sign}}
\def\be{\begin{eqnarray}}
\def\ee{\end{eqnarray}}
\newcommand{\nn}{\nonumber\\}
\newcommand{\Eq}[1]{Eq.~(\ref{#1})}

\newcommand{\cR}{ {\cal R} }

\begin{document}

\title{Checkerboard charge density wave and pseudogap in high-$T_{c}$
 cuprates}

\author{Jian-Xin Li}
\affiliation{Department of Physics, University of California at
Berkeley, Berkeley, CA 94720, USA} \affiliation{National
Laboratory of Solid State Microstructure, Nanjing University,
Nanjing 210093, China}

\author{Chang-Qin Wu}
\affiliation{Department of Physics, University of California at
Berkeley, Berkeley, CA 94720, USA} \affiliation{Department of
Physics, Fudan University, Shanghai 200433, China}

\author{Dung-Hai Lee}
\affiliation{Department of Physics, University of California at
Berkeley, Berkeley, CA 94720, USA} \affiliation{Material Science
Division, Lawrence Berkeley National Laboratory, Berkeley, CA
94720, USA}

\date{\today}

\begin{abstract}
We consider the scenario where a 4-lattice constant, rotationally
symmetric charge density wave (CDW) is present in the underdoped
cuprates. We prove a theorem that puts strong constraint on the
possible form factor of such a CDW.  We demonstrate, within
mean-field theory, that a particular form factor within the allowed
class describes the angle-resolved photoemission and scan tunneling
spectroscopy well. We conjecture that the ``large pseudogap'' in
cuprates is the consequence of this type of charge density wave.
\end{abstract}

\maketitle

\section{Introduction}

After almost two decades of experimental study, it is known that
the high temperature superconductors have the following known
ordered states: 1) antiferromagnetic order at very low doping
($x\lesssim 3\%$), 2) the $d$-wave superconducting (DSC) order for
$5\%\lesssim x\lesssim 30\%$. While these two orders exist in all
families of cuprates, there is a third order, namely, 3) a
4-lattice constant charge and 8-lattice constant spin density wave
order, occurring near doping $x=1/8$ in the
La$_{1.48}$Nd$_{0.4}$Sr$_{0.12}$CuO$_4$/La$_{1.875}$Ba$_{0.125}$CuO$_4$
(LNSCO/LBCO) systems\cite{tranquada}. There is a wide-spread
belief that this charge/spin density wave order is anisotropic,
i.e., they form stripes\cite{tranquada,zannen,ek,rez}.

A significant part of the high-$T_c$ mystery lies in the behavior of
the underdoped systems\cite{palee}. Based on specific
heat\cite{loram}, nuclear magnetic resonance \cite{NMR} , DC
transport\cite{transport}, optical and Raman
spectroscopy\cite{puchkov,blum}, angle-resolved photoemission
(ARPES) \cite{damascelli} and tunneling
\cite{fisher,lang,mcelroy,hanaguri,kohsaka} , Tallon and Loram have
made the case that the high-$T_c$ superconductors possess two energy
gaps, a pseudogap and a superconducting gap\cite{loram}. Recently
ARPES experiments on LSCO systems\cite{fujimori} and underdoped
Bi2212\cite{zxshen} both point to a large pseudogap in the antinodal
region and a superconducting gap near the Brillouin zone diagonals.
Similar result has also been found in electronic Raman scattering
experiment on Hg1201\cite{tacon}. In addition, it is shown that for
underdoped Bi2212 a large pseudogap exists in the antinodal region
even at temperature $\approx 3 T_c$, while a gapless Fermi arc
exists near the nodes.\cite{kenigel}

Recently, there are clear evidences from the the scan tunneling
spectroscopy (STM) studies suggesting the presence of a 4-lattice
constant checkerboard order in NaCCOC\cite{hanaguri} and
underdoped Bi2212\cite{mcelroy,kohsaka}. Interestingly ARPES study
has shown that in Na$_x$Ca$_{2-x}$CuO$_2$Cl$_2$, where STM found
checkerboard order\cite{hanaguri}, the Fermi arcs
survive\cite{kshen}.

In view of these new experimental results we ask the question
``can the pseudogap in underdoped cuprates be caused by some kind
of checkerboard CDW?''. To answer the question, we will look at
the effects of the checkerboard CDW on {\it low energy}
quasiparticles. Since the existence of low energy quasiparticles
is an experimental fact, it is reasonable to model the influence
of CDW by an {\it effective} scattering Hamiltonian of the form
\be H_{CDW}=\sum_{\v Q}\sum_{\v k}\sum_{\sigma} [ f(\v {k,Q})
C^+_{\v k+\v Q\sigma}C_{\v k\sigma} + h.c.],\label{cdw}\ee where
$\v Q$ is the CDW ordering wavevector and $f(\v {k,Q})$ the form
factor.

In the following, we will first explore the symmetry property of
the checkerboard CDW form factor using the experimentally observed
STM patterns in Sec.II. In Sec.III, we compare the low energy
ARPES and STM spectral functions generated by two representatives
among the allowed form factors. Section IV is the summary.

\section{Two theorems about $f(\v k,\v Q)$}

\begin{figure}
\includegraphics[angle=270,scale=0.5]{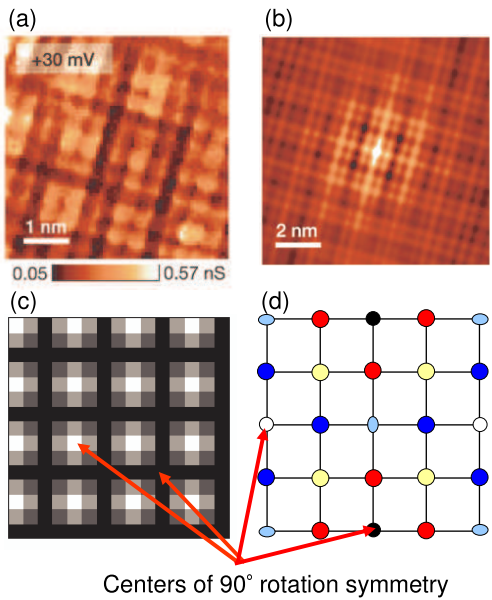}
\caption{\label{fig1} (Color online) STM dI/dV map (a) and the
autocorrelation image of $|E|<100$meV LDOS maps (b) from Hanaguri
{\it et al.}~\cite{hanaguri} on NaCCOC, showing the $4\times 4$
ordering. (c) Caricature of the observed image shown in (a). (d)
Possible LDOS pattern which exhibits 6 independent intensities in
the $4\times 4$ unit cell. In panels (c) and (d), two
nonequivalent $s$-symmetry centers are indicated by arrows. In
panel (d), the $d$-symmetry centers are indicated by the ellipses.
}
\end{figure}
In Fig.~\ref{fig1}(a), we reproduce the  STM dI/dV image of
Na$_x$Ca$_{2-x}$CuO$_2$Cl$_2$ from Ref.14. This particular image
is made at bias voltage 30 mV. However, the same checkerboard
pattern was seen in a wide bias range $-150 {\rm mV}\le V\le 150
{\rm mV}$. Experimentally, it was determined that such a
checkerboard pattern contains $\pm \v Q$, where \be\v Q=
(2\pi/4,0),(0, 2\pi/4)\label{Q}\ee as its fundamental ordering
wavevector. Hence we limit the $\v Q$ summation in \Eq{cdw} to
those given by \Eq{Q} and {\v k} to the first Brillouin zone. In
Fig.~\ref{fig1}(b), we reproduce the two-point correlation
function of the observed image presented in Ref.14. From
Fig.1(a,b) we construct a caricature in Fig.~\ref{fig1}(c) to
capture the essence of the observed checkerboard. Interestingly,
in each $4\times 4$ unit cell there are two inequivalent centers
about which the checkerboard is symmetric under\be
C_{4v}=\{E,C_2,\sigma_x,\sigma_y,C_4,C_4^3,\sigma_{x+y},\sigma_{x-y}\},\ee
the point group of the square lattice. (Here $E$ represents
identity, and $C_{2,4}$ denote 180 and 90 degree rotations, and
$\sigma$ denotes reflection.) In the following, we take this as
implying that
$H_{\rm CDW}$ is $C_{4v}$-invariant about these two centers.\\

 {\bf
Theorem I} A CDW that has the ordering wavevectors given by \Eq{Q}
and possesses a center of $C_{4v}$ symmetry in
its unit cell must have the following properties. \\
\indent    (1) There must exist another inequivalent $C_{4v}$ center
in the unit cell. This second center is displaced from the first by
the (2,2)
translation or its equivalent. About these two centers $f(\v k,\v Q)$ has $s$-symmetry. \\
\indent    (2) There must exist two other centers around which
$H_{\rm CDW}$ remains invariant under $C_{2v}$, the subgroup formed
by the first four elements of $C_{4v}$, but changes sign under
$C_4,C_4^3,\sigma_{x+y},\sigma_{x-y}$. Spatially these two new
centers must be displaced from the two $C_{4v}$ centers by the (2,0)
and (0,2) translation or their equivalents. About these two centers $f(\v k,\v Q)$ has $d$-symmetry.\\

{\it Proof.} Let us assume $H_{\rm CDW}$ is invariant under
$C_{4v}$ at the origin, i.e., ${\cR}H_{\rm CDW}{\cR}^{-1}=H_{\rm
CDW}$ where $\cR\in C_{4v}$. This implies \be f(\cR\v k,\cR\v
Q)=f({\v {k, Q}})\ee from Eq.(1). After a translation {\v t} the
form factor changes to \be f({\v {k, Q}})\rightarrow g({\v {k,
Q}})\equiv f({\v {k, Q}})e^{i\v{Q\cdot t}},\ee where {\v t} can be
any one of the 16 possible displacements within the unit cell \be
{\v t}=(m,n),~~m,n=0,1,2,3.\ee Due to the fact that {\v Q}  only
takes one of the four possible values given in \Eq{Q} it can be
easily checked that for $\v t =(2,2)$ \be g(\cR\v k,\cR\v Q)=g({\v
{k, Q}})~~\forall\cR\in C_{4v},\ee and for $ {\v t}=(2,0), (0,2)$
\be &&g(\cR\v k,\cR\v Q)=g({\v {k, Q}})~~\cR\in C_{2v}\nn&&g(\cR\v
k,\cR\v Q)=-g({\v {k, Q}})~~\cR\in C_{4v}-C_{2v},\ee where
$C_{4v}-C_{2v}\equiv\{
C_4,C_4^3,\sigma_{x+y},\sigma_{x-y}\}$. {\it QED}\\

Theorem I implies that any 90-degree rotationally symmetric CDW
with $(\pm 2\pi/4,0),(0,\pm 2\pi/4)$ ordering wavevectors must
{\it simultaneously possess s-symmetry centers} {\it and
d-symmetry centers}. The presence of both symmetry centers is a
necessary consequence of the CDW being rotationally symmetric.
This fact was overlooked in the earlier version of this paper.
Conversely any four lattice constant CDW that does not possess
both symmetry centers must break rotation symmetry.  In addition,
it can be shown easily that a rotationally symmetric CDW discussed
above possesses 6 inequivalent sites in the unit cell~\cite{hu},
hence allowing 6 different
values of dI/dV. This is shown in Fig.~\ref{fig1}(d).\\

{\bf Theorem II} If $H_{\rm CDW}$ is time reversal invariant, $f(\v
{k,Q})$ must be real if one chooses either $d$- or $s$-symmetry
center as the origin.\\

 {\it Proof.} Time reversal symmetry requires
\be f^*({\v {k,Q}})=f({\v {-k,-Q}}).\ee Since $f({\v {k,Q}})$ is
invariant under the 180 degree rotation about the $s$ and $d$
centers we have \be f({\v {k,Q}})=f({\v {-k,-Q}}).\ee As a result,
\be f^*({\v {k,Q}})=f({\v {k,Q}}),\ee i.e, $f({\v {k,Q}})$ is
real. $QED$

\section{Effects of the CDW on ARPES and STM spectra}

In this section we apply the two theorems proven above and take
the input from a previous renormalization group
calculation\cite{fu} to guess the plausible form of $f(\v k,\v
Q)$. We then investigate the effect of the checkerboard CDW on the
STM and ARPES spectral functions of the low energy quasiparticles.
We stress that the purpose of this section is
not to prove that the ground state of certain microscopic Hamiltonian has CDW order. 
Rather, we take a phenomenological approach by {\it assuming} its
existence and look at its consequences that are observable by STM
and ARPES.

In Ref.23 it was shown that, with the help of electron-phonon
interaction, a class of electron-electron scattering is enhanced
at low energies. This class of scattering involves (momentum
conserving) scattering of a pair of quasiparticles near the
antinodes. For example, consider a pair of quasiparticles lying on
the opposite sides of the almost nested Fermi surface near the
$(\pi,0)$ antinodes as shown in Fig.2(a). After the scattering
these two quasiparticles switch sides. The momentum transfer in
such a scattering is the ``nesting wavevector'' of the antinodes.
For systems such as NaCCOC\cite{kshen} and underdoped
Bi2212\cite{feng} it has been shown that such nesting wavevectors
are approximately given by \Eq{Q}. Interestingly, Ref.23 also
shows that accompanying each such scattering there is a related
process, whose scattering amplitude has opposite sign, where one
of the quasiparticle scattering takes place near the $(0,\pi)$
rather than the $(\pi,0)$ antinode [Fig.2(b)]. It was also noticed
that when this type of quasiparticle scattering grows strong it
tends to drive a CDW whose form factor has the property that \be
\sign[f(\cR\v k,\v Q)]=-\sign[f(\v k,\v Q)]~{\rm for}~\cR\in
C_{4v}-C_{2v}.\label{fu}\ee

In the following, let us choose the $d$-symmetry center as the
origin. Thus \be f(\cR\v k,\cR\v Q)=-f(\v k,\v Q)~{\rm for}~\cR\in
C_{4v}-C_{2v}.\label{d}\ee Combine \Eq{d} with \Eq{fu} we obtain
\be \sign[f(\v k,\cR\v Q)]=\sign[f(\v k,\v Q)]~{\rm for}~\cR\in
C_{4v}-C_{2v}.\label{fu1}\ee In addition, \Eq{fu} plus the
continuity condition requires \be f(\v k,\v Q)=0~~{\rm for}~~\v k
~~{\rm along }~~ \hat{x}\pm\hat{y}.\ee The above considerations
lead us to the following ansaz for the CDW form factor \be f(\v
k,\v Q)=S_{\v k}(\v Q)(\cos k_x-\cos k_y)\equiv S_{\v k}(\v
Q)f_0(\v k),\label{ans}\ee where $S_{\v k}(\v Q)>0$. In the
following we shall pick a simple realization of \Eq{ans} and focus
on $\v k$ lying close to the Fermi surface.

In general the CDW couples each $\v k$ to other 15 $\v k$ points in
the first Brillouin zone. However, most of these 16 $\v k$'s lie far
away from the Fermi surface, hence can be omitted in the low-energy
theory. This suggests that one only needs to keep a few close
neighbors for each $\v k$. Another important consideration guiding
our construction of $H_{\rm CDW}$ is the requirement that {\it a
robust antinodal CDW gap exists for reasonable change of doping.} It
turns out that this requirement is satisfied as long as the nested
scattering across the antinodal Fermi surface is the dominant
scattering process.

Put all the constraints together we consider the following
quasiparticle Hamiltonian in the absence of superconducting
pairing\be\mathcal{H}=\sum_{\v k,\sigma} \Psi_{\sigma}^+(\v k)
\mathcal{A}(\v k) \Psi_{\sigma}(\v k),\label{ch}\ee where
\be\Psi_{\sigma}^+(\v k)=( c^+_{\v k,\sigma},c^+_{\v k+\v
Q_1,\sigma},c^+_{\v k+\v Q_2,\sigma},c^+_{\v k-\v Q_2,\sigma}),\ee
and \be
 \mathcal{A}(\v k)=\pmatrix{\epsilon_{\v k}&S_0f_0(\v k)&S_1f_0(\v k)&S_2f_0(\v k)\cr
S_0f_0(\v k)&\epsilon_{\v k+\v Q_1}&0&0\cr S_1f_0(\v
k)&0&\epsilon_{\v k+\v Q_2}&0\cr S_2f_0(\v k)&0&0&\epsilon_{\v
k-\v Q_2}}.\label{cdwh}\ee In \Eq{cdwh} \be &&\v
Q_1=-\sign(k_x)(2\pi/4,0), \v Q_2=(0,2\pi/4)~{\rm for~}
|k_x|<|k_y|\nn&&\v Q_1=-\sign(k_y)(0,2\pi/4), \v Q_2=(2\pi/4,0)
~{\rm for~} |k_x|>|k_y|\nonumber\ee as shown schematically in
Fig.2(c). In addition, we expect $S_0$ to be stronger than $S_1$
and $S_2$.
\begin{figure}
\includegraphics[scale=0.85]{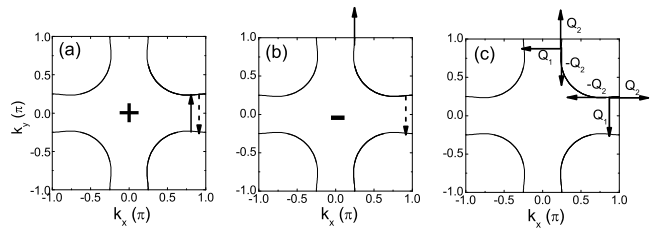}
\caption{\label{fig:epsart} The two enhanced sets of
electron-electron scattering (panels (a) and (b)), as obtained
from a renormalization group calculation~\cite{fu}. The scattering
amplitude between these two sets differs by a sign. (c) The
CDW-induced quasiparticle scatterings (only those in the first
quadrant of the Brillouin zone are shown). The solid lines in
these figures represent the normal state Fermi surface.}
\end{figure}
For the normal state dispersion, we use $\epsilon_{\v
k}=t_{0}+t_{1}[\cos(k_{x}) + \cos(k_{y})]/2 +
t_{2}\cos(k_{x})\cos(k_{y})+t_{3}[\cos(2k_{x}) + \cos(2k_{y})]/2
+t_{4}[\cos(2k_{x})\cos(k_{y}) + \cos(2k_{y})\cos(k_{x})]/2 +
t_{5}\cos(2k_{x})\cos(2k_{y})$, with the hopping constants (in eV)
$(t_1,...,t_{5})=(-0.5951,0.1636,-0.0519,-0.1117,0.0510)$\cite{norman}.
In the following, we will compare the effects of the CDW for the
two cases where the Fermi surface is nested/not nested by the $\v
Q$ given by \Eq{Q}. (We adjust $t_0$ to control the degree of
nesting.) As to the CDW order parameter, we choose \be
&&S_0=\Delta_c,~~S_1=s\Delta_c,~~S_2=s\Delta_c
.\label{ff}\ee
\begin{figure*}
\includegraphics[scale=1.6]{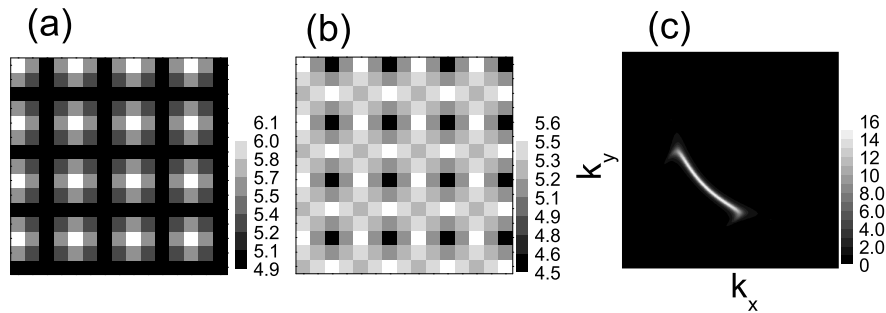}
\caption{\label{fig:epsart}  Panel (a) and (b) are the dI/dV images
for the CDW state where the $f_0(\v k)$ in \Eq{ch} and \Eq{cdwh} is
$\cos k_{x}-\cos k_{y}$ and $|\cos k_{x}-\cos k_{y}|$, respectively.
The window of view is $15\times 15$ lattice unit cells. Panel (c) is
their Fermi energy ARPES intensity maps (Both form factors give the
same intensity map). Here only the first quadrant of the Brillouin
zone is shown. In making these figures we have chosen $\Delta_c$ to
produce a 60 meV gap at the antinodes. The parameter $s$ in \Eq{ff}
is chosen to be 0.2. A quasiparticle energy broadening of 10 meV and
a $t_{0}=0.0945$ eV are used.}
\end{figure*}

We first discuss the case with Fermi surface nesting. In Fig.3(a)
and (c) we present the real space dI/dV image at bias voltage $20$
mV and the ARPES intensity map at the Fermi level. These results
are calculated with $s=0.2$ in \Eq{ff}. The primary effect of
changing $s$ is to 1) change the intensity variation in the black
perimeter in each unit cell in Fig.3(a); and 2) affect the
strength of shadow band in Fig.3(c) (see later). Except these
changes, the main features of both results are preserved. In
Fig.3(b) we show the dI/dV image resulting from \Eq{ch} where the
$f_0(\v k)$ in \Eq{cdwh} is replaced by  $|\cos k_x-\cos k_y|$ (Of
course, after such a choice the $d$-symmetry center becomes the
$s$-symmetry center). The purpose of this figure is to demonstrate
the sensitivity of the real space image on the sign of $f_0$.
Indeed, while the ARPES image is completely unaffected by such a
change, the real space dI/dV is strongly modified. Upon a
comparison with the checkerboard pattern observed in
Na$_x$Ca$_{2-x}$CuO$_2$Cl$_2$\cite{hanaguri}, it is clear that the
form factor $\cos k_{x}-\cos k_{y}$ (Fig.3(a)) produces the real
space description best. To better understand the Fermi arc present
in Fig.3(c) we note that in the presence of CDW, the new Fermi
surface is determined by \be det[A(\v k)]=0.\ee Since $det[A(\v
k)]$ is real (because $A(\v k)$ is Hermitian) $det[A(\v k)]=0$
yields a single equation with two unkowns ($k_x$ and $k_y$).
Generically, one expects the solutions to form {\it closed}
one-dimensional curves. Since $f_0(\v k)$ vanishes at the node, it
is natural to expect the Fermi surface to be practically
unaffected in its vicinity. Such an unaffected piece of the Fermi
surface and its CDW shadows form a closed contour. The reason that
in Fig.3(c) only a Fermi arc is visible is due to the CDW
coherence factor\cite{chak}. In Fig.3(c), the strongest shadow
band effect shows up near the end of the Fermi arcs. Note that
such shadow band position is very different from that expected
from antiferromagnetism. Presently there is no report of seeing
such shadow bands\cite{chatterjee, kenigel}. The reason may be: 1)
the CDW correlation length as observed by STM experiment is not
sufficiently long (It is typically of 10 nanometers); 2) in the
pseudogap regime, the superconducting pairing still persists. In
all cases we studied, the superconducting pairing is very
effective in weakening the shadow band effect. When moving away
from the zero binding energy, we find that the main changes in the
ARPES intensity map are: 1) the intensity in the antinodal regions
increases, and 2) the Fermi arcs shrink and move towards the
origin of the first Brillouin zone.

By considering all panels of Fig.3, it is obvious that it is the
checkerboard CDW with  $f_0(\v k)=\cos k_{x}-\cos k_{y}$ that
reproduces both the ARPES and STM phenomenology well. Therefore, we
will only consider this kind of form factor in the rest of the
paper.

\begin{figure*}
\includegraphics[scale=1.8]{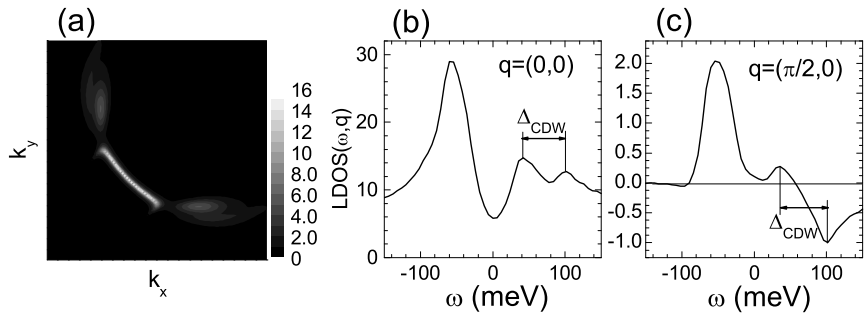}
\caption{\label{fig:epsart} (a) The ARPES intensity at $E_F$ for
the checkerboard CDW with the form factor $\cos k_{x}-\cos k_{y}$
and a 60 meV gap. (b) and (c) are the $\v q=(0,0)$ and $\v
q=(\pi/2,0)$ Fourier components of LDOS for a state with a 60 meV
CDW gap and a $\Delta_{0}=40$ meV DSC pairing parameter. A
$t_{0}=0.1215$ eV is used.}
\end{figure*}
Now, we turn to the case without Fermi surface nesting. In this
case, using the checkerboard CDW with an order parameter of the
same magnitude as that in Fig.3, we obtain a weaker fragmentation
of the Fermi surface as shown in Fig.4(a). As to the real space
pattern (not shown), the only difference with Fig.3(b) is a slight
increase in the intensity variation in the dark perimeter region.

Next, we turn on a DSC pairing and ask what is the signature of the
checkerboard CDW and superconducting pairing coexistence in STM. In
this case the Hamiltonian becomes \be\mathcal{H}=\sum_{\v k}
\Phi^+(\v k) \mathcal{H}(\v k) \Phi(\v k),\ee where \be\Phi^+(\v
k)=\pmatrix{\Psi^+_{\uparrow}(\v k),\Psi_{\downarrow}(-\v k)},\ee
and \be \mathcal{H}(\textbf{k})=\left(\begin{array}{cc} A(\textbf{k}) & B(\textbf{k})\\
B^*(\textbf{k}) & -A(\textbf{-k})\end{array}\right).\label{tew}\ee
In the above equations  \be &&B_{ij}(\v k)=0~{\rm for}~ i\ne j\nn&&
B_{ii}(\v k)=\Delta_{\v k},\Delta_{\v k+\v Q_1},\Delta_{\v k+\v
Q_2},\Delta_{\v k-\v Q_2}~{\rm for}~ i=1,2,3,4.\nn\ee For d-wave
superconducting (DSC) pairing $\Delta_{\v k}=\Delta_{0}(\cos
k_{x}-\cos k_{y})/2$. In the presence of inversion symmetry ($A(-\v
k)=A(\v k)$) the Hamiltonian in \Eq{tew} can also be written as
\be\mathcal{H}(\textbf{k})=A(\textbf{k})\otimes
\sigma_3+B(\textbf{k})\otimes\sigma_1.\ee In that case because
  $\mathcal{H}(\textbf{k})$ anticommutes with $I\otimes\sigma_y$ the
eigen spectrum is particle-hole symmetric. Under such condition the
zero-energy eigenvectors are also eigenvectors of
$I\otimes\sigma_y$. As a result, the locus of zero energy satisfies
\be det[A(\v k)\pm i B(\v k)]=0.\ee Since this determinant is
complex, setting its real and imaginary parts to zero gives two
equations for the two unknown $k_x$ and $k_y$. Consequently, one
expects the solutions to be isolated points in the Brillouin zone.
Thus with the DSC pairing the Fermi arc produced by checkerboard CDW
is reduced to point gap nodes.

In Fig.4(b) and (c) we consider the case where a $60$ meV
checkerboard CDW order parameter coexists with a $\Delta_{0}=40$
meV DSC pairing. Fig.4(b) shows the spatial averaged local density
of states (LDOS). Note that the CDW feature on the negative bias
side is much weaker than that of the positive side. This is
because it is overwhelmed by the density of states due to the van
Hove singularity. The two peaks on the positive bias side are the
original antinodal coherence peak split by the CDW order. We have
checked that the energy separation between these peaks is
proportional to the CDW order parameter. Another way to determine
the strength of the CDW order is to Fourier transform LDOS at the
CDW ordering wavevector. In Fig.4(c), the real part of the $\v
q=(\pi/2,0)$ component of LDOS is shown. The two peaks on the
positive bias side of Fig.4(b) now appear as a peak and an
anti-peak. Again, the distance between them is proportional to the
CDW order parameter. Thus we propose that by studying the Fourier
transformed LDOS, it is possible to extract the strength of CDW
ordering.

In Fig.5(a), we show several ARPES momentum distribution curves
(MDC) along the momentum cut $(-\pi/2,\pi)\rightarrow (\pi/2,\pi)$
for the checkerboard CDW. All energies considered here are below
the CDW gap. The presence of two non-dispersive MDC peaks
separated by the CDW ordering wavevector is apparent. This is very
similar to that observed in Ref.21.

In Fig.5(b), we present the energy gap along the normal state
Fermi surface for a pure checkerboard CDW state (dashed curve) and
a state with both checkerboard CDW and DSC pairing (solid curve).
The purpose of this figure is to illustrate the effect of DSC
pairing in the pseudogap state. It shows how Fermi arc is replaced
by a gap node. With thermal phase fluctuations, this explains why
Fermi arcs shrink to four points as temperature approaches zero as
observed recently\cite{kenigel}. Given these results, we feel
quite tempted to associate the larger checkerboard CDW gap with
the large pseudogap and the smaller pairing gap on the Fermi arc
with the small pseudogap.

\begin{figure}
\includegraphics[scale=1.2]{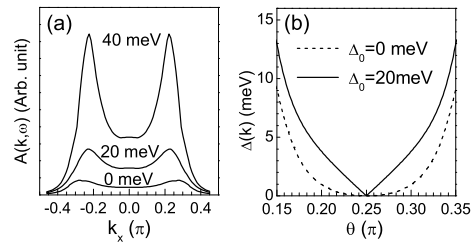}
\caption{\label{fig:epsart} (a) The ARPES MDC along the momentum
cut $(-\pi/2,\pi)\rightarrow (\pi/2,\pi)$ at three different
energies in the presence of a checkerboard CDW state. (b) The
energy gap along the normal state Fermi surface for the pure
checkerboard CDW state (dashed curve) and a state with coexisting
CDW and DSC order (solid curve). The form factor of the CDW is
$\cos k_{x}-\cos k_{y}$ and the CDW parameters are the same as
those in Fig.3. }
\end{figure}

In the literature it is widely believed that the pseudogap is a
consequence of the short-range antiferromagnetic
correlation\cite{palee}. Thus it is natural to ask what is the
relation between the checkerboard CDW discussed above and such
physics. On microscopic level the CDW presented in this paper
represents the modulation in the hopping (or antiferromagnetic
exchange) integrals. Consequently, it is a kind of spin Peirls
distortion which, of course, is compatible with the spin singlet
pairing tendency of a quantum antiferromagnet. In addition to the
above remarks we note that in a recent paper\cite{wzq} it is found
that checkerboard CDW is a self-consistent solution of a $t-J$ like
model at mean-field level, again testify that checkerboard CDW does
not contradict the superexchange physics.

\section{Conclusion}

In this paper, we present a symmetry constraint on the form factor
of a 90 degree rotationally symmetric, commensurate, checkerboard
charge density wave. Further guided by a previous renormalization
group study\cite{fu} we construct a simple model describing the
scattering of the low energy quasiparticles by the CDW.  We then
calculate the low energy ARPES and STM spectra using this simple
model. The results compare favorably with the existing
experiments. In particular, the results show a spatial dI/dV
pattern similar to the one observed in
Na$_x$Ca$_{2-x}$CuO$_2$Cl$_2$ and underdoped Bi2212 by
STM\cite{hanaguri,kohsaka}. Moreover, in the momentum space it
produces Fermi arcs resembling those observed by
ARPES\cite{fujimori,kshen}. In the presence of a $d$-wave
superconducting pairing, the Fermi arcs of the checkerboard CDW
are reduced to four gap nodes\cite{kenigel}. Therefore, this study
supports the notion that the large antinodal pseudogap in
underdoped cuprates is generated by the checkerboard charge
density wave\cite{ddw,cdw,cdw1} conjectured at the beginning of
the paper.

\section{acknowledgement}
We thank Seamus Davis, Henry Fu, Gey-Hong Gweon, Alessandra
Lanzara, and Zhi-Xun Shen for valuable discussions. JXL and CQW
acknowledge the support of the Berkeley Scholar program and the
NSF of China. DHL was supported by the Directior, Office of
Science, Office of Basic Energy Sciences, Materials Sciences and
Engineering Division, of the U.S. Department of Energy under
Contract No. DE-AC02-05CH11231.

\end{document}